\documentclass[12pt]{article}
\usepackage{mathrsfs}
\usepackage{amssymb}
\usepackage{epsfig}
\normalsize
\input{epsf}

\begin{document}
\addtolength{\baselineskip}{.20mm}
\newlength{\extraspace}
\setlength{\extraspace}{2mm}
\newlength{\extraspaces}
\setlength{\extraspaces}{2mm}

\newcommand{\newsection}[1]{
\vspace{15mm} \pagebreak[3] \addtocounter{section}{1}
\setcounter{subsection}{0} \setcounter{footnote}{0}
\noindent {\Large\bf \thesection. #1} \nopagebreak
\medskip
\nopagebreak}

\newcommand{\newsubsection}[1]{
\vspace{1cm} \pagebreak[3] \addtocounter{subsection}{1}
\addcontentsline{toc}{subsection}{\protect
\numberline{\arabic{section}.\arabic{subsection}}{#1}}
\noindent{\large\bf 
\thesubsection. #1} \nopagebreak \vspace{3mm} \nopagebreak}
\newcommand{\ba}{\begin{eqnarray}
\addtolength{\abovedisplayskip}{\extraspaces}
\addtolength{\belowdisplayskip}{\extraspaces}

\addtolength{\belowdisplayshortskip}{\extraspace}}

\newcommand{\be}{\begin{equation}
\addtolength{\abovedisplayskip}{\extraspaces}
\addtolength{\belowdisplayskip}{\extraspaces}
\addtolength{\abovedisplayshortskip}{\extraspace}
\addtolength{\belowdisplayshortskip}{\extraspace}}
\newcommand{\ee}{\end{equation}}
\newcommand{\STr}{{\rm STr}}
\newcommand{\figuur}[3]{
\begin{figure}[t]\begin{center}
\leavevmode\hbox{\epsfxsize=#2 \epsffile{#1.eps}}\\[3mm]
\parbox{15.5cm}{\small
\it #3}
\end{center}
\end{figure}}
\newcommand{\im}{{\rm Im}}
\newcommand{\calm}{{\cal M}}
\newcommand{\call}{{\cal L}}
\newcommand{\sect}[1]{\section{#1}}
\newcommand\hi{{\rm i}}
\def\bea{\begin{eqnarray}}
\def\eea{\end{eqnarray}}

\begin{titlepage}
\begin{center}

\vspace{3.5cm}

{\Large \bf{Noether Symmetry Approach in  multiple scalar fields Scenario}}\\[1.5cm]

{Yi Zhang $^{a,b,}$\footnote{Email: zhangyia@cqupt.edu.cn},}{Yun-Gui
Gong $^{a}$\footnote{Email: gongyg@cqupt.edu.cn}, }{Zong-Hong Zhu
$^{b,}$\footnote{Email: zhuzh@bnu.edu.cn},} \vspace*{0.5cm}

{\it $^{a}$College of Mathematics and Physics,
Chongqing Universe of Posts and Telecommunications, \\ Chongqing 400065, China

$^{b}$ Department of Astronomy, Beijing Normal University,
\\  Beijing 100875, China}

\date{\today}
\vspace{3.5cm}

\textbf{Abstract} \vspace{5mm}

\end{center}
In this Letter, we find  suitable potentials in the multiple scalar
fields scenario by using the Noether symmetry approach. We discussed
three models with multiple scalar fields: N-quintessence with
positive kinetic terms, N-phantom with  negative kinetic terms and
N-quintom with both positive and negative kinetic terms. In the
N-quintessence case, the exponential potential which could be
derived from several theoretic models is obtained from the Noether
conditions. In the N-phantom case, the potential
$\frac{V_{0}}{2}(1-\cos(\sqrt{\frac{3N}{2}}\frac{\phi}{m_{pl}}))$,
which could be derived from the Pseudo Nambu-Goldstone boson model,
is chosen as the Noether conditions required. In the N-quintom case,
we derive a relation  $DV'_{\phi q}=-\tilde{D}V'_{\phi p}$ between
the potential forms for the quintessence-like fields and the
phantom-like fields by using the Noether symmetry.
\end{titlepage}

\section{Introduction}\label{sec1}
Scalar field theory which is related to particle physics has become the generic playground for building cosmological models,
both in  the early and  late accelerating  periods of our universe \cite{Guth:1980zm,Riess:1998cb}.
Although the dynamics of these accelerations is likely to contain several scalar fields,
it is normally assumed that only one of these fields
remained dynamically significant for a long time. However,  realistic theoretical models, embedded in grand unified
or super symmetric theories, must necessarily be
 theories of multiple fields.
The simplest multiple scalar fields scenario which we will consider is first originated from the assisted inflation
scenario \cite{Liddle:1998jc}.
The essential point of this scenario is that inflation is not driven by any single field, but a
collection of $N$ fields. These fields have the same initial conditions and potentials.
This idea can be applied in vector field models  as well \cite{Golovnev:2008cf}.

Meanwhile, the observations suggest the equation of state (EoS)
parameter of dark energy is in the range of $-1.21\leq \omega\leq
-0.89$ \cite{Riess:2004nr}.  Since the quintessence type of matter
could not give the possibility that $\omega<-1$, the extended
paradigms (e.g. phantom and quintom) are proposed. Phantom type of
matter with  negative kinetic energy  has well-known problems, but,
nevertheless, was  implicitly suggested
 in cosmological models and
 have also been widely
studied as  dark energy. It is phenomenologically significant
 and worthy of putting other theoretical difficulties aside temporally. Then, it is natural to ask why don't we discuss the multiple scalar fields with different kinetic terms.
According to the classification of
 the scalar fields\footnote{ The quintessence with positive kinetic term was proposed in Ref.\cite{Ratra:1987rm};
the phantom with negative kinetic term was suggested in Ref. \cite{Caldwell:1999ew}; and
 the quintom  with both positive and  negative kinetic terms was proposed in Ref. \cite{Feng:2004ad}.},
 we can  discuss three types of fields in the simplest multiple scalar fields scenario, which are the quintessence type of fields with positive kinetic terms,
 the phantom type of fields with
 negative kinetic terms, the quintom type of fields with both positive and  negative kinetic terms.
 In this Letter, we call them N-quintessence, N-phantom, N-quintom for convenience.

But, as in the single scalar field
 case, we have to ask how to choose the  potentials from the
various models for those multiple scalar fields.  In this Letter, we
will deal with this problem of choice from a point of view of
symmetry. The Noether symmetry has been revealed as a useful tool
for finding out exact solutions in cosmology. This is an
 interesting method to select models motivated at a fundamental level.

 This Letter is organized as follows. In section \ref{sec2},
we introduce the multiple scalar fields models.
 In section  \ref{sec3}, the  Noether symmetry  approach will be introduced and applied
to both N-quintessence and N-phantom cases to get exact solutions.
In section  \ref{sec4}, we discuss the application of Noether
symmetry approach to N-quintom  case in connection with its
solution. In section  \ref{addsec}, we give out the evolution of our
universe in  N-quintessence and N-phantom cases. Finally, a short
summary will be presented in section \ref{sec5}.


\section{N-quintessence, N-phantom and N-quintom Scalar Field   Model Scenario}\label{sec2}
 As stated in the introduction, usually, only
 one scalar field is enough to accelerate the universe, but a single field  is not natural. The application of the multiple scalar fields in cosmology should be seriously considered.
 Here, we assume that the geometry of space-time is described by the flat FRW (Friedmann-Robertson-Walker) metric which seems to be
consistent with today's cosmological observations
\begin{eqnarray}
 ds^{2}=-dt^{2}+a^{2}(t)\sum^{3}_{i=1}(dx^{i})^{2},
\end{eqnarray}
where $a$ is the scale factor.
After setting the number of the scalar fields as $N$, the action of the multiple scalar fields can be written as
\begin{eqnarray}
 \label{action}
 S_{\phi}=\int d^{4}x \sqrt{-g}\left[\frac{R}{16 \pi G}+\sum_{i=1}^{N}\left(\epsilon\frac{\dot{\phi_{i}}^{2}}{2}-V(\phi_{i})\right)\right],
\end{eqnarray}
where $\epsilon=1$ denotes the quintessence fields with the positive kinetic terms,
$\epsilon=-1$ denotes the phantom fields
with the negative kinetic term. Meanwhile, as we consider both the vector fields and the matter in the system, the total action  is
 \be
 S_{tot}=S_{\phi}+S_{m},
 \ee
 where $S_{m}$ is the action for matter. The density of the matter can be expressed as $\rho_{m}=\rho_{m0}(a_{0}/a)^{3\gamma}$, where $\rho_{m0}$ is an initial constant and $0<\gamma\leq2$.
Here, we limit our analysis to $\gamma=1$ which corresponds to the pressureless matter with $P_{m}=0$.

We assume the vector fields are non-interacting,
their influences on each other are through their effects on the expansion. Considering
  all the scalar fields have the same potentials
 and initial conditions,
 action (\ref{action}) could be simplified as
\begin{eqnarray}
   \label{action1}
 S_{\phi1}=\int d^{4}x \sqrt{-g}\left[\frac{R}{16 \pi G}+N\left(\epsilon\frac{\dot{\phi}^{2}}{2}-V(\phi)\right)\right].
\end{eqnarray}
When $\epsilon=1$, we call the related scenario  N-quintessence.
While $\epsilon=-1$, we call the related scenario N-phantom.

For the N-quintom case, we assume the fields with same kinetic terms
have the same potentials and initial conditions,
the action  can be written as
\begin{eqnarray}
   \label{action2}
 S_{\phi2}=\int d^{4}x \sqrt{-g}\left[\frac{R}{16 \pi G}+N_{q}\left(\frac{\dot{\phi_{q}}^{2}}{2}
 -V(\phi_{q})\right)+N_{p}\left(-\frac{\dot{\phi_{p}}^{2}}{2}-V(\phi_{p})\right)\right],
\end{eqnarray}
where $\phi_{q}$ is the scalar field with the positive kinetic terms,  $N_{q}$ is the number of the corresponding quintessence type fields;
 $\phi_{p}$ is the scalar field with the negative kinetic terms,
 $N_{p}$ is the number of the corresponding phantom type fields.
This paradigm  has been  proved of crossing $\omega_{\phi}=-1$ when $N_{q}=N_{p}=1$ \cite{Feng:2004ad}.

\section{The Noether Symmetry Approach in N-quintessence and N-phantom}\label{sec3}
 In the case of N-quintessence and N-phantom, we take the scale factor $a$ and the scalar
field $\phi$ as independent dynamical variables in the  system which the action
(\ref{action1})  represents.
Then  the  configuration space could be chosen as  $\mathcal{Q}=(a,\phi)$, while   the related tangent space is $T\mathcal{Q}=(a,\phi,\dot{a},\dot{\phi})$.
To study the symmetries of the  space under consideration, we
need an effective point-like Lagrangian for the model whose variation with respect to
its dynamical variables yields the correct equations of motion.  However, based on  action (\ref{action1}), it is proper to make the point-like Lagrangian as
\begin{eqnarray}
\label{L} \mathcal{L}_{1}=\mathcal{L}_{\phi1}+\mathcal{L}_{m}=
3a\dot{a}^{2}-\frac{N}{m_{pl}^{2}}\left(\epsilon\frac{a^{3}\dot{\phi}^{2}}{2}-a^{3}V(\phi)\right)+\frac{\rho_{m0}}{m_{pl}^{2}},
\end{eqnarray}
where the Planck mass is $m_{pl}^{2}=(8\pi G)^{-1}$, and the term $\frac{\rho_{m0}}{m_{pl}^{2}}$ corresponds to the effects from matter.

Therefore,  the total energy of the system  $E_{\mathcal{L}_{\phi1}}$,  could be written in this way
\begin{eqnarray}
\label{E}
E_{\mathcal{L}_{1}}=\frac{\partial\mathcal{L}_{1}}{\partial
\dot{q_{i}}}\dot{q_{i}}-\mathcal{L}_{1}=a^{3}\left(\frac{\epsilon
N\dot{\phi}^{2}}{2}+NV(\phi)+\rho_{m0}a^{-3}-3m_{pl}^{2}H^{2}\right).
\end{eqnarray}
If the above equation being considered as a constraint, with the
vanishing of the ``energy function" , it is just the Friedmann
equation
 \be \label{fr1}
H^{2}=\frac{1}{3m_{pl}^{2}}\left[\frac{\epsilon N
\dot{\phi}^{2}}{2}+NV(\phi)+\rho_{m0}a^{-3}\right]. \ee

Furthermore, for a dynamical system, the Euler-Lagrangian equation is
\begin{eqnarray}
\frac{d}{dt}(\frac{\partial\mathcal{L}_{1}}{\partial
\dot{q_{i}}})-\frac{\partial\mathcal{L}_{1}}{\partial q_{i}}=0.
\end{eqnarray}
Based on the Lagrangian, in the N-quintessence and N-phantom case,
the variable $q_{i}$ is $a$ and $\phi$, respectively.
When $q_{i}=a$,  the Raychaudhuri equation could be gotten
 \be
 \label{me1}
\dot{H}=-\frac{m_{pl}^{2}}{2}(\rho_{\phi}+P_{\phi}+\rho_{m})=-\frac{\epsilon Nm_{pl}^{2}}{2}\dot{\phi}^{2}-\frac{m_{pl}^{2}}{2}\rho_{m},
 \ee
 where the energy density and the pressure of scalar fields are
\begin{eqnarray}
&&\rho_{\phi}=\frac{\epsilon N}{2}\dot{\phi}^{2}+NV(\phi),\\
&&P_{\phi}=\frac{\epsilon N}{2}\dot{\phi}^{2}-NV(\phi).
\end{eqnarray}
What is more, the equation of state could also be obtained \be
\omega_{\phi}=\frac{P_{\phi}}{\rho_{\phi}}=\frac{\epsilon
\dot{\phi}^{2}/2-V(\phi)}{\epsilon \dot{\phi}^{2}/2+V(\phi)}. \ee
Obviously, in the N-quintessence case where $\epsilon=1$,
$\omega_{\phi}>-1$; in the N-phantom case where $\epsilon=-1$,
$\omega_{\phi}<-1$. Both of them could not cross $\omega_{\phi}=-1$,
that is why we also consider N-quintom. In the case of $q_{i}=\phi$,
the Euler-Lagrangian equation is the equation of motion
 \be
 \label{me}
\ddot{\phi}+3H\dot{\phi}+\epsilon V'_{\phi}=0, \ee where the prime
means $V'_{\phi}=dV/d\phi$. For the different value of $\epsilon$,
the quintessence makes the fields roll down the potential, while the
phantom makes them roll up.

The above equations coincide with the results calculated from the Einstein equations,
and prove that the point-like Lagrangian is consistent with the dynamical system.

As is well known in \cite{deRitis:1990ba,various model,f(R)}, Noether symmetry
approach is a powerful tool in finding the solution for a given Lagrangian.
From this method, it is possible to obtain a reduction, and possibly get a full integration of the system, whenever the cyclic variable of the system  is found.
 The key point related to the Noether symmetry is a Lie algebra  presented in the tangent space. Following \cite{deRitis:1990ba,various model,f(R)},  for the Lagrangian (\ref{L}),  firstly
 we  define the Noether symmetry induced  by a vector $X$  on the tangent space $T\mathcal{Q}=(a,\phi,\dot{a},\dot{\phi})$
 which is
\begin{eqnarray}
\label{X}
X=\alpha\frac{\partial}{\partial a}+\beta\frac{\partial}{\partial \phi}+\dot{\alpha}\frac{\partial}
{\partial \dot{a}}+\dot{\beta}\frac{\partial}{\partial \dot{\phi}},
\end{eqnarray}
 where $\alpha$ and $\beta$ are generic functions of $a$ and $\phi$.
The Lagrangian is invariant under the transformation $X$ if
\begin{eqnarray}
\label{L1}
L_{X}\mathcal{L}_{1}=\alpha\frac{\partial\mathcal{L}_{1}}{\partial
a}+\frac{d \alpha}{dt}\frac{\partial \mathcal{L}_{1}}{\partial
\dot{a}}+\beta\frac{\partial\mathcal{L}_{1}}{\partial a}+\frac{d
\beta}{dt}\frac{\partial \mathcal{L}_{1}}{\partial \dot{a}}=0.
\end{eqnarray}
 Given $L_{X}\mathcal{L}_{1}=0$  satisfied, there  exists a  Noether symmetry.
Combined with the Lagrangian, this symmetry  gives out
\begin{eqnarray}
&&\alpha+2a\frac{\partial \alpha}{\partial a}=0,\\
\label{2}
&&6\frac{\partial \alpha}{\partial \phi}-\epsilon N\frac{a^{2}}{m_{pl}^{2}}\frac{\partial \beta}{\partial a}=0,\\
\label{3}
&&3\alpha+2 a\frac{\partial \beta}{\partial \phi}=0,\\
\label{4} &&3V(\phi)\alpha+aV'_{\phi}(\phi)\beta=0,
\end{eqnarray}
which we call Noether conditions.
The difference between the N-quintessence and N-phantom is in  Eq.(\ref{2}) as  the parameter $\epsilon$ denotes.

What is more, the momentum potential  can be defined as below
\begin{eqnarray}
\label{pa}
&&p_{a}=\frac{\partial\mathcal{L}_{1}}{\partial \dot{a}}=6a\dot{a},\\
\label{pphi} &&p_{\phi}=\frac{\partial \mathcal{L}_{1}}{\partial
\dot{\phi}}=-\frac{\epsilon N}{m_{pl}^{2}}a^{3}\dot{\phi}.
\end{eqnarray}
Then we can express the constant of motion which is reproduced by the Noether symmetry
\begin{eqnarray}
\label{Q}
\alpha p_{a}+\beta p_{\phi}=Q=\mu_{0},
\end{eqnarray}
where $Q$ is called conserved charge and $\mu_{0}$ is the related constant.
 The Noether constant of motion on shell gives a possibility of
 solving the system. More specifically, a symmetry exists if at least one of the
functions $\alpha$ or $\beta$ is different from zero. As a byproduct, the form of $V(\phi)$ is determined in correspondence with such a symmetry.

 The cyclic variable can be regarded as a helpful tool of  getting the   exact description about the dynamical system.
A point transformation $(a,\phi)\rightarrow(z,w)$ is effective to find the cyclic variable. It is
\begin{eqnarray}
\label{z}
&&i_{X}z=\alpha\frac{\partial z}{\partial \dot{a}}+\beta\frac{\partial z}{\partial \dot{\phi}}=1,\\
\label{w}
&&i_{X}w=\alpha\frac{\partial w}{\partial \dot{a}}+\beta\frac{\partial w}{\partial \dot{\phi}}=0,
\end{eqnarray}
then the Lagrangian could  be rewritten in term of the cyclic variables. After the transformation,
 the cyclic variable is $z$, and the constant of motion can be rewritten as $Q=p_{z}$. This will simplify our calculation effectively.
A general discussion of this issue could be found in \cite{deRitis:1990ba,various model,f(R)}.
After introducing the Noether symmetry approach,  we will discuss the solutions for the Noether conditions both in the N-quintessence and N-phantom in the following.

\subsection{Exact solutions for N-quintessence}\label{subsec1}

In  the N-quintessence case where the sign of the kinetic terms takes the value $\epsilon=1$,  the Noether conditions are
\begin{eqnarray}
&&\alpha+2a\frac{\partial \alpha}{\partial a}=0,\\
\label{q2}
&&6\frac{\partial \alpha}{\partial \phi}- N\frac{a^{2}}{m_{pl}^{2}}\frac{\partial \beta}{\partial a}=0,\\
&&3\alpha+2 a\frac{\partial \beta}{\partial \phi}=0,\\
&&3\alpha V(\phi)+a\beta V'_{\phi}(\phi)=0.
\end{eqnarray}
When $N=1$, the Noether conditions reduce to the   single field case \cite{various model}.  As indicated by Eq.(\ref{q2}),
the effects of the multiple scalar fields are manifested by the number of the scalar fields $N$.

An obvious constant potential solution is
\be
\alpha=0,\,\, \beta={\rm constant},\,\, V={\rm constant}.
\ee
In this solution, $a$ is the cyclic variable.
And the subsequent constant of motion gives out
\be
\beta p_{\phi}=-\frac{\epsilon N}{m_{pl}^{2}}a^{3}\dot{\phi}=Q=\mu_{0}.
\ee
The discussions could be divided into two cases simply. Firstly,
when $\mu_{0}=0$, $\phi=constant$, this is a cosmological constant solution.
Secondly, when $\mu_{0}\neq0$, the kinetic term $\dot{\phi}\propto a^{3}$.
The scalar fields decay fast, even faster than the corresponding  vector field solution \cite{zhang}.
These two cases are trivial respectively.  In the following, we will concentrate our discussions on  another solution which is
\begin{eqnarray}
\label{alpha}
&&\alpha=\frac{\sigma_{+}}{\sqrt{a}},\,\,\,\beta=\frac{-3\lambda \sigma_{-}}{2a\sqrt{a}},\\
\label{V}
&&V=V_{0}\sigma_{-}^{2}=V_{0}(A^{2}e^{2\lambda\phi}+B^{2}e^{-2\lambda \phi}-2AB),
\end{eqnarray}
where $\sigma_{\pm}=A e^{\lambda \phi}\pm B e^{-\lambda \phi}$,  $\lambda=\sqrt{3N/8m_{pl}^{2}}$, $A$ and $B$ are constants.

We can see that the potential is a combined exponential function. Indeed,
there are some physical origins about this kind of potential.
In higher-dimensional gravitational theories such as superstring and Kaluza-Klein theories \cite{expmoti},
exponential potentials often appear from the curvature of internal spaces associated with the geometry of extra dimensions \cite{gaugino}. Moreover,
it is known that exponential potential can arise in gaugino condensation as a non-perturbative effect and in the presence of supergravity
corrections to global supersymmetric theories \cite{CNR}. However, this kind of potential is picked up by Noether symmetry.

In particular, when $A=0$, the Noether conditions show
\be
V(\phi)=V_{0}\exp{\left(-\sqrt{\frac{3N}{2}}\frac{\phi}{m_{pl}}\right)}.
\ee
This kind of potential leads to a power-law expanding universe, with
$a\propto t^{4/3}$, $\omega_{\phi}=-1/2$. The quintessence with an exponential potential
was widely studied in cosmology, see, for example, Ref. \cite{Ferreira}.
It even has a scaling solution.
In the following, based on the value of $A$,   we will get the exact solutions
from the point of view of Noether symmetry.

\subsubsection{when $A\neq0$ and $B\neq0$}\label{subsubsec1}
If we put Eqs. (\ref{pa}), (\ref{pphi}), (\ref{alpha}) and  (\ref{V}) into  Eq. (\ref{Q}),
we find that the constant of motion is hard to obtain.
Therefore, we search  the cyclic variable for help.
By calculating Eqs. (\ref{z}) and (\ref{w}), we can get the following expressions for the new variables
\begin{eqnarray}
z=\frac{a^{3/2}\sigma_{+}}{6AB},\,\, w=\frac{a^{3/2}\sigma_{-}}{6AB},
\end{eqnarray}
where $z$ is the cyclic variable.
Correspondingly,   $\phi$ and $a$ could be expressed as
\begin{eqnarray}
\label{35}
\phi=\frac{1}{2\lambda}\ln \frac{z+w}{z-w},\,\,
a=\left[9AB(z^{2}-w^{2})\right]^{1/3}.
\end{eqnarray}
The resulting forms of  potential and  Lagrangian are
\begin{eqnarray}
&&V(\phi)=V_{0}\frac{4w^{2}}{z^{2}-w^{2}},\\
&&\mathcal{L}_{\phi1}=12AB\left[(\dot{z}^{2}-\dot{w}^{2})+\frac{3NV_{0}}{m_{pl}^{2}}w^{2}\right].
\end{eqnarray}
Using the Euler-Lagrangian equations, the above Lagrangian leads to the
equations of motion for $z$ and $w$,
\begin{eqnarray}
\ddot{z}=0,\,\,\ddot{w}=-\frac{3NV_{0}}{m_{pl}^{2}}w.
\end{eqnarray}
The solutions are
\begin{eqnarray}
&&z=z_{1}t+z_{0},\\
&&w=w_{1}\sin(\sqrt{\frac{3NV_{0}}{m_{pl}^{2}}}t+w_{0}),
\end{eqnarray}
where $z_{0}$, $z_{1}$, $w_{0}$, $w_{1}$ are constants.
Therefore, the exact evolution of the field and the scale factor  could be given out as below
\begin{eqnarray}
&&\phi=\frac{1}{2\lambda}\ln \frac{z_{1}t+z_{0}+w_{1}\sin(\sqrt{\frac{3NV_{0}}{m_{pl}^{2}}}t+w_{0})}{z_{1}t+z_{0}-[w_{1}\sin(\sqrt{\frac{3NV_{0}}{m_{pl}^{2}}}t+w_{0})]},\\
&&a=\left[9AB((z_{1}t+z_{0})^{2}-w_{1}^{2}\sin^{2}(\sqrt{\frac{3NV_{0}}{m_{pl}^{2}}}t+w_{0}))\right]^{1/3}.
\end{eqnarray}
If $z\ll w$, we could not get a physical value of $\phi$, through  the scale factor seems oscillate.
And if  $z \gg w$, $\phi$ is very small, but the universe will evolve as $a\propto t^{2/3}$. It is similar to the matter-dominated phase.

\subsubsection{when $A=0$ and $B\neq0$}\label{subsubsec2}
 In this subsection, we continue to search the cyclic variables but for a different potential where  $A=0$ while $B\neq0$.
By calculating Eqs. (\ref{z}) and (\ref{w}),  the expressions of the new variables are
\begin{eqnarray}
z=\frac{a^{3/2}}{3\sigma_{+}},\,\,w=\frac{a^{3/2}}{3\sigma_{+}},
\end{eqnarray}
where $z$ is the cyclic variable.
Then   $\phi$ and $a$ can be rewritten as
\begin{eqnarray}
\label{44}
\phi=\frac{1}{2\lambda}\ln (\frac{B^{2}zw}{9}),\,\,a=(\frac{z}{w})^{1/3}.
\end{eqnarray}
As a result, we  get the potential and the Lagrangian in term of $z$ and $w$
\begin{eqnarray}
&&V(\phi)=\frac{V_{0}}{9zw},\\
&&\mathcal{L}_{\phi1}=\frac{-4}{3}\frac{\dot{z}\dot{w}}{w^{2}}+\frac{N}{m_{pl}^{2}}\frac{V_{0}}{9w^{2}}.
\end{eqnarray}
Apply the new Lagrangian to the Euler-Lagrangian equations,  we obtain
\begin{eqnarray}
\ddot{z}=\frac{3V_{0}}{2m_{pl}^{2}w},\,\,
\ddot{w}=\frac{2\dot{w}^{2}}{w}.
\end{eqnarray}
They lead to
\begin{eqnarray}
&&z=-\left[\frac{V_{0}w_{2}}{4m_{pl}^{2}}t^{3}+\frac{3V_{0}w_{3}}{4m_{pl}^{2}}t^{2}+\frac{3V_{0}w_{4}}{4m_{pl}^{2}}t+w_{5} \right],\\
&&w=\frac{-1}{w_{2}t+w_{3}},
\end{eqnarray}
where $w_{2}$, $w_{3}$, $w_{4}$ are constants. Putting the above equations into Eq. (44),
the evolutions of  $a$ and $\phi$ are
\begin{eqnarray}
&&\phi=\sqrt{\frac{2}{3N}}m_{pl} \ln
\left(B^{2}\frac{\frac{V_{0}w_{2}}{4m_{pl}^{2}}t^{3}+\frac{3V_{0}w_{3}}{4m_{pl}^{2}}t^{2}+
\frac{3V_{0}w_{4}}{4m_{pl}^{2}}t+w_{5}}{w_{2}t+w_{3}}\right),\\
&&a=\left[(w_{2}t+w_{3})(\frac{V_{0}w_{2}}{4m_{pl}^{2}}t^{3}+\frac{3V_{0}w_{3}}{4m_{pl}^{2}}t^{2}+\frac{3V_{0}w_{4}}{4m_{pl}^{2}}t+w_{5}) \right]^{1/3}.
\end{eqnarray}
When $z\varpropto t$, the scale factor evolves as  $a\propto
t^{2/3}$ which is similar to  the matter-dominated phase. When
$z\varpropto t^{3}$, the scale factor is $a\propto t^{4/3}$ which
may accelerate the universe. This is an interesting solution that we
need. We will discuss this solution in section (\ref{addsec}) in
detail.

However, the N-quintessence scenario could be replaced by  a  single field paradigm with the similar evolutions   $a\propto t^{4/3}$.
 We just need to change the corresponding  parameter in the single field case as
 \be
 V_{0s}=NV_{0}, \,\,\lambda_{s}=\frac{\lambda}{\sqrt{N}}.
 \ee
The reason
for this behavior is that  each field experiences the `downhill' force from its own potential, it feels the
friction from all the scalar fields via their contribution to the expansion rate.

The case $B=0$, $A\neq0$ is treated exactly in the same way and the results are the same, except for the  substitution of $A$ for $B$.
In summary, it must be noted that our results include some already known models.
The exponential potential not only make the acceleration last a long time, but also satisfy the Noether conditions.

As for the comparison with the observations, one field results have been derived by Ref. \cite{deRitis:1990ba}. In the N-quintessence case, the range of parameter will be changed because of $N$. Considering our purpose is on the choice of the potential, we will not discuss this subject in detail.

\subsection{Exact solutions for N-Phantom}\label{subsec2}
For the N-phantom case where  $\epsilon=-1$, the Noether  conditions are
\begin{eqnarray}
&&\alpha+2a\frac{\partial \alpha}{\partial a}=0,\\
\label{p2}
&&6\frac{\partial \alpha}{\partial \phi}+ N\frac{a^{2}}{m_{pl}^{2}}\frac{\partial \beta}{\partial a}=0,\\
&&3\alpha+2 a\frac{\partial \beta}{\partial \phi}=0,\\
&&3\alpha V(\phi)+a\beta V'_{phi}(\phi)=0.
\end{eqnarray}
Compared to the N-quintessence case, the difference arises in Eq. (\ref{p2})  by the sign of the  kinetic terms.

Obviously, the simplest solution  is
\be
\alpha=0,\,\,\beta={\rm constant},\,\,V={\rm constant}.
\ee
This constant potential solution is similar to the corresponding solution in the N-quintessence case,
we don't discuss this fast decaying case.

However, another  interesting solution is
\begin{eqnarray}
&&\alpha=\frac{2C\cos(\frac{1}{2}\sqrt{\frac{3N}{2}}\frac{\phi}{m_{pl}})}{\sqrt{a}},\\
&&\beta=\frac{-2\sqrt{6}C\sin(\sqrt{\frac{-3N}{8}}\frac{\phi}{m_{pl}})}{a\sqrt{a}},\\
\label{V2}
&&V(\phi)=V_{0}\sin^{2}(\frac{1}{2}\sqrt{\frac{3N}{2}}\phi)=\frac{V_{0}}{2}(1-\cos(\sqrt{\frac{3N}{2}}\frac{\phi}{m_{pl}})),
\end{eqnarray}
where $C$ is a constant.
When $N=1$, there are some differences between the results in  Ref. \cite{Capozziello:2009te} and ours.
The form of the potential could be
called PNGB (Pseudo Nambu-Goldstone Bosons) potential resulting from explicit breaking of a shift symmetry \cite{Freese:1990rb}.

To find the exact evolution of the universe,  as the calculations in
the N-quintessence case, we need the help of  the cyclic variables.
 According to Eqs. (\ref{z}) and (\ref{w}), a transformation could be done from $(a,\phi)$ to $(z,w)$,
\be
\phi=\arctan\frac{w}{z},\,\,\,a=(3C)^{2/3}(z^{2}+w^{2})^{1/3},
\ee
then we can rewrite the potential and the Lagrangian as
\begin{eqnarray}
&&V=\frac{V_{0}w^{2}}{z^{2}+w^{2}},\\
&&\mathcal{L}_{\phi1}=9C^{2}\left[
\frac{4}{3}(\dot{z}^{2}+\dot{w}^{2})+V_{0}\frac{Nw^{2}}{m_{pl}^{2}}\right].
\end{eqnarray}
The Lagrangian leads to the equations of motion for the new variables
\begin{eqnarray}
\ddot{z}=0,\,\,\ddot{w}=\frac{3}{4}\frac{NV_{0}w}{m_{pl}^{2}}.
\end{eqnarray}
The solutions are
\begin{eqnarray}
&&z=z_{3}t+z_{2},\\
&&w=w_{6}\exp(\sqrt{\frac{3}{4}\frac{NV_{0}}{m_{pl}^{2}}}t),
\end{eqnarray}
where $z_{2}$, $z_{3}$ and $w_{6}$ are constant.
However, by using the cyclic variable $z$, we  get the evolutions of the field and the scale factor,
\begin{eqnarray}
&&\phi=\arctan\frac{w_{6}\exp(\sqrt{\frac{3}{4}\frac{NV_{0}}{m_{pl}^{2}}}t)}{z_{3}t+z_{2}},\\
\label{a3} &&a=(3C)^{2/3}\left[
(z_{3}t+z_{2})^{2}+w_{6}^{2}\exp(\sqrt{\frac{3NV_{0}}{m_{pl}^{2}}}t)\right]^{1/3}.
\end{eqnarray}
If $z\gg w$, the values of fields are nearly zero,
 $a\propto t^{2/3}$, it is the matter-dominated solution.
When $z\ll w$, the universe evolves as
$a\propto\exp(\sqrt{\frac{3NV_{0}}{m_{pl}^{2}}}t)$, this is the
de-Sitter solution. We will discuss this solution in section
(\ref{addsec}) in detail.

As for the comparison with the observations, one field results have been
derived by Ref. \cite{Capozziello:2009te}. In the N-phantom case, the range
of parameter will be changed because of $N$. Considering our purpose is on the choice of the potential, we will not discuss this subject in detail.

\section{Noether Symmetry in N-Quintom Case}\label{sec4}
The quintom scenario is proposed to fit the observable data \cite{Riess:2004nr}.
N-quintessence and N-phantom could not cross $\omega_{\phi}=-1$ as we see.
However, N-quintom has  an  attractive feature that it may cross $\omega_{\phi}=-1$ which is a possibility  implied by the data.
After adding the Noether symmetry, this property should be rechecked.
Though in the ``cosmic triad" vector field  case, Noether symmetry provides an interesting constraint on
 the potentials \cite{zhang} for the quintom case with $\omega_{\phi}$ crossing $-1$. However, it is worthy of trying the Noether symmetry approach in the N-quintom case.
According to the action (\ref{action2}),
the point-like Lagrangian  is
\begin{eqnarray}
\label{L2} \mathcal{L}_{2}=
3a\dot{a}^{2}-\frac{N_{q}}{m_{pl}^{2}}\left(\frac{a^{3}\dot{\phi_{q}}^{2}}{2}-a^{3}V_{q}\right)-\frac{N_{p}}
{m_{pl}^{2}}\left(\frac{-a^{3}\dot{\phi_{p}}^{2}}{2}-a^{3}V_{p}\right)+\rho_{m0}.
\end{eqnarray}

Based on the above point-like Lagrangian, the total energy and the Euler-Lagrangian equation will give out
the Friedmann equation, the Raychaudhuri equation and the equations of motion
\begin{eqnarray}
 &&H^{2}=\frac{1}{3m_{pl}^{2}}\left[N_{q}(\frac{ \phi_{q}^{2}}{2}+V_{q})+N_{p}(-\frac{ \phi_{p}^{2}}{2}+V_{p})+\rho_{m}\right],\\
 \label{meq}
&&\dot{H}=-\frac{m_{pl}^{2}}{2}(\rho_{\phi}+P_{\phi}+\rho_{m})=-\frac{m_{pl}^{2}N_{q}}{2}\dot{\phi}_{q}^{2}+\frac{m_{pl}^{2}N_{p}}{2}\dot{\phi}_{p}^{2}-\frac{m_{pl}^{2}}{2}\rho_{m},\\
&&\ddot{\phi}_{q}+3H\dot{\phi}_{q}+ V'_{\phi q}=0,\\
&&\ddot{\phi}_{p}+3H\dot{\phi}_{p}- V_{p}'=0,
\end{eqnarray}
where the primes mean $V'_{\phi q}=dV_{q}/d\phi_{q}$ and $V'_{\phi
p}=dV_{p}/d\phi_{p}$. The energy density and the pressure which
could be derived from the action (\ref{action2}) are
\begin{eqnarray}
&&\rho_{\phi}=N_{q}(\frac{\dot{\phi_{q}}^{2}}{2}+V_{q})+N_{p}(-\frac{\dot{\phi_{p}}^{2}}{2}+V_{p}),\\
&&P_{\phi}=N_{q}(\frac{\dot{\phi_{q}}^{2}}{2}-V_{q})-N_{p}(\frac{\dot{\phi_{p}}^{2}}{2}+V_{p}).\\
\end{eqnarray}
So the EoS parameter is
\be
\label{w4}
\omega_{\phi}=\frac{N_{q}(\frac{\dot{\phi_{q}}^{2}}{2}-V_{q})-N_{p}(\frac{\dot{\phi_{p}}^{2}}{2}+V_{p})}{N_{q}(\frac{\dot{\phi_{q}}^{2}}{2}+V_{q})+N_{p}(-\frac{\dot{\phi_{p}}^{2}}{2}+V_{p})}.
\ee

Now, we should choose  a new configuration space $\mathcal{Q}=(a,\phi_{q},\phi_{p})$ with
the corresponding tangent space $T\mathcal{Q}=(a,\phi_{q},\phi_{p},\dot{a},\dot{\phi_{q}},\dot{\phi_{p}})$.
And the vector generator which induce the Noether symmetry is changed to
\begin{eqnarray}
\label{X}
\tilde{X}=\tilde{\alpha}\frac{\partial}{\partial a}+\tilde{\beta}\frac{\partial}{\partial \phi_{q}}+\gamma\frac{\partial}{\partial \phi_{p}}+
\dot{\tilde{\alpha}}\frac{\partial}{\partial \dot{a}}+\dot{\tilde{\beta}}\frac{\partial}{\partial \dot{\phi_{q}}}+\dot{\gamma}\frac{\partial}{\partial \dot{\phi_{p}}},
\end{eqnarray}
where $\tilde{\alpha}$, $\tilde{\beta}$ and $\gamma$ are generic
functions of the variables $a$, $\phi_{q}$ and $\phi_{p}$. The
Noether symmetry requires the Lie derivative  of the Lagrangian
vanishes  which means $L_{\tilde{X}}\mathcal{L}_{2}=0$. Following
Ref. \cite{deRitis:1990ba,various model,f(R)}, the Noether
conditions can be obtained
\begin{eqnarray}
&&\tilde{\alpha}+2a\frac{\partial \tilde{\alpha}}{\partial a}=0,\\
&&6\frac{\partial \tilde{\alpha}}{\partial \phi_{q}}- \frac{N_{q}a^{2}}{m_{pl}^{2}}\frac{\partial \tilde{\beta}}{\partial a}=0,\\
&&6\frac{\partial \tilde{\alpha}}{\partial \phi_{p}}+ \frac{N_{p}a^{2}}{m_{pl}^{2}}\frac{\partial \gamma}{\partial a}=0,\\
&&3\tilde{\alpha}+2 a\frac{\partial \tilde{\beta}}{\partial \phi_{q}}=0,\\
&&3\tilde{\alpha}+2 a\frac{\partial \gamma}{\partial \phi_{p}}=0,\\
&&3(V_{q}+V_{p})\tilde{\alpha}+aV'_{\phi q}\tilde{\beta}+aV'_{\phi
p}\gamma=0.
\end{eqnarray}

There is an obvious  solution that is
 \be
 \tilde{\alpha}=0,\,\, \tilde{\beta}=D,\,\, \gamma=\tilde{D},
 \ee
 where $D$ and $\tilde{D}$ are integral constants.
The symmetry  exists, if and only if at least one of the parameter
$\tilde{\alpha}$, $\tilde{\beta}$, $\gamma$ is not zero. Based on
the Noether conditions, we find a condition relating the potential
forms of the quintessence-like fields and the phantom-like fields,
that is \be \label{V3} DV'_{\phi q}=-\tilde{D}V'_{\phi p}. \ee And
the constant of motion corresponding to this solution  is \be
\label{Q4} -D N_{q} a^{3}\dot{\phi}_{q}+\tilde{D}N_{p}
a^{3}\dot{\phi}_{p}=Q=\mu_{0}. \ee

In the following discussion, based on the value of $D$, $\tilde{D}$ and $\mu_{0}$, we  try to discuss the solutions, especially for the value of EoS parameter.

\subsection{When $D\neq0$ and $\tilde{D}=0$ }
 If  $D\neq0$ and $\tilde{D}=0$, we can get  $V'_{\phi q}=0$, the quintessence-like matter has a constant potential.
And from the constant of motion, we can get  $\dot{\phi_{q}}^{2}\propto \mu_{0}^{2}a^{-6}$.
 However, based on the value of $\mu_{0}$, we divide the situation into two cases to discuss.

Case a), when  $\mu_{0}\neq 0$,
the kinetic terms of the quintessence decay fast, while their potentials are constant, and no constraint on the phantom type of matter, which  leads to $w<-1$ at last.

Case b), when $\mu_{0}= 0$,
the quintessence scalar filed is a constant. This case is similar to a phantom model  with cosmological constant.
The interesting thing is that we could not give any constraint on the phantom-like matters.

 The case $D=0$ and $\tilde{D}\neq0$ could be treated exactly in the same way.
 And the results are the same, except for the non-constrained field is changed to the quintessence-like type.

\subsection{When  $D\neq0$ and    $\tilde{D}\neq0$}

\subsubsection{The $\mu_{0}\neq 0$ case}
In this case, the conserved charge is not zero. From Eq. (\ref{Q4}),
we get that  $\dot{\phi}_{q}=D\dot{\phi}_{p}/\tilde{D}\propto
a^{-3}$. It means that the kinetic terms of the scalar field decay
fast. The equations of motion leads to $V'_{\phi p}=V'_{\phi q}=0$,
i.e., the potentials are constant. However,
 the EoS parameter evolves to $\omega_{\phi}=-1$ until the kinetic terms of the scalar fields vanish.

\subsubsection{The $\mu_{0}=0$ case}\label{sub4.2.2}
In this case, the conserved charge vanishes, so $D N_{q} \dot{\phi}_{q}=\tilde{D} N_{p} \dot{\phi}_{p}$,
combined  with Eq. (\ref{V3}) and the equations of motion,  $N_{q}=N_{p}$ is obtained.
We put these results into Eq. (\ref{w4}), and get
\be
\omega_{\phi}=\frac{\frac{(1-D^{2}/\tilde{D}^{2})\dot{\phi_{q}}^{2}}{2}-V_{q}-V_{p}}{\frac{ (1-D^{2}/\tilde{D}^{2})\dot{\phi_{q}}^{2}}{2}+V_{q}+V_{p}}.
\ee
If $D/\tilde{D}<1$, $\dot{\phi_{p}}^{2}<\dot{\phi}_{q}^{2}$, $\omega_{\phi}>-1$.
The physical meaning  is that if the quintessence type fields slowly vary  compared with the phantom
 type fields,  the quintessence will take the dominating role, and make $\omega_{\phi}>-1$.
And we can discuss the $D/\tilde{D}>1$ case in the same way, where  the phantom type fields will take the dominating role and $\omega_{\phi}<-1$.
However,
this solution is  new.  And if it cross $\omega_{\phi}=-1$,
 the ratio $D/\tilde{D}$ should be variable.
However, as Noether symmetry approach required, $D/\tilde{D}$ is constant.
It means in N-quintom case, after adding Noether symmetry, we could not make this scenario cross $\omega_{\phi}=-1$.

In a short summary, even the Noether symmetry does not give an explicit potential in N-quintom case,
it gives a constraint on the forms of
the scalar field potentials. If we try to connect this model to the observations such as SNIa data, we must choose a proper potential. Unfortunately, the observations will give constraints to the potential parameter not the parameter related to Noether symmetry which we are interested here. And this symmetry restricts the EoS parameter of  crossing  $\omega_{\phi}=-1$.

\section{From Deceleration to Acceleration}\label{addsec}

\begin{figure}[ht]
\begin{center}
\includegraphics[width=15cm]{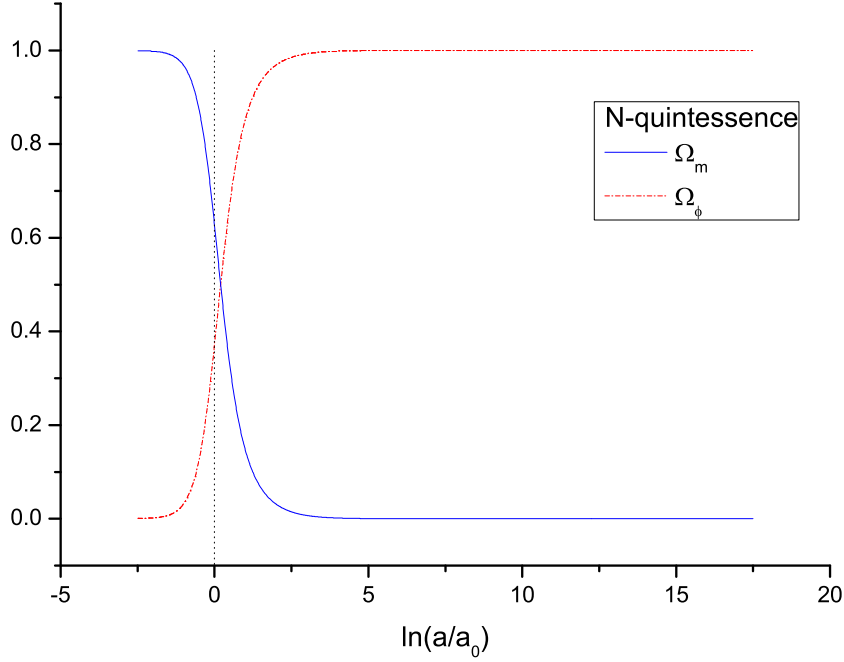}
\caption{\label{fig1}The evolutions of fractional energy densities
$\Omega_\phi$ and $\Omega_{m}$ in N-quintessence model.}
\label{fig:68}
\end{center}
\end{figure}



\begin{figure}[ht]
\begin{center}
\includegraphics[width=15cm]{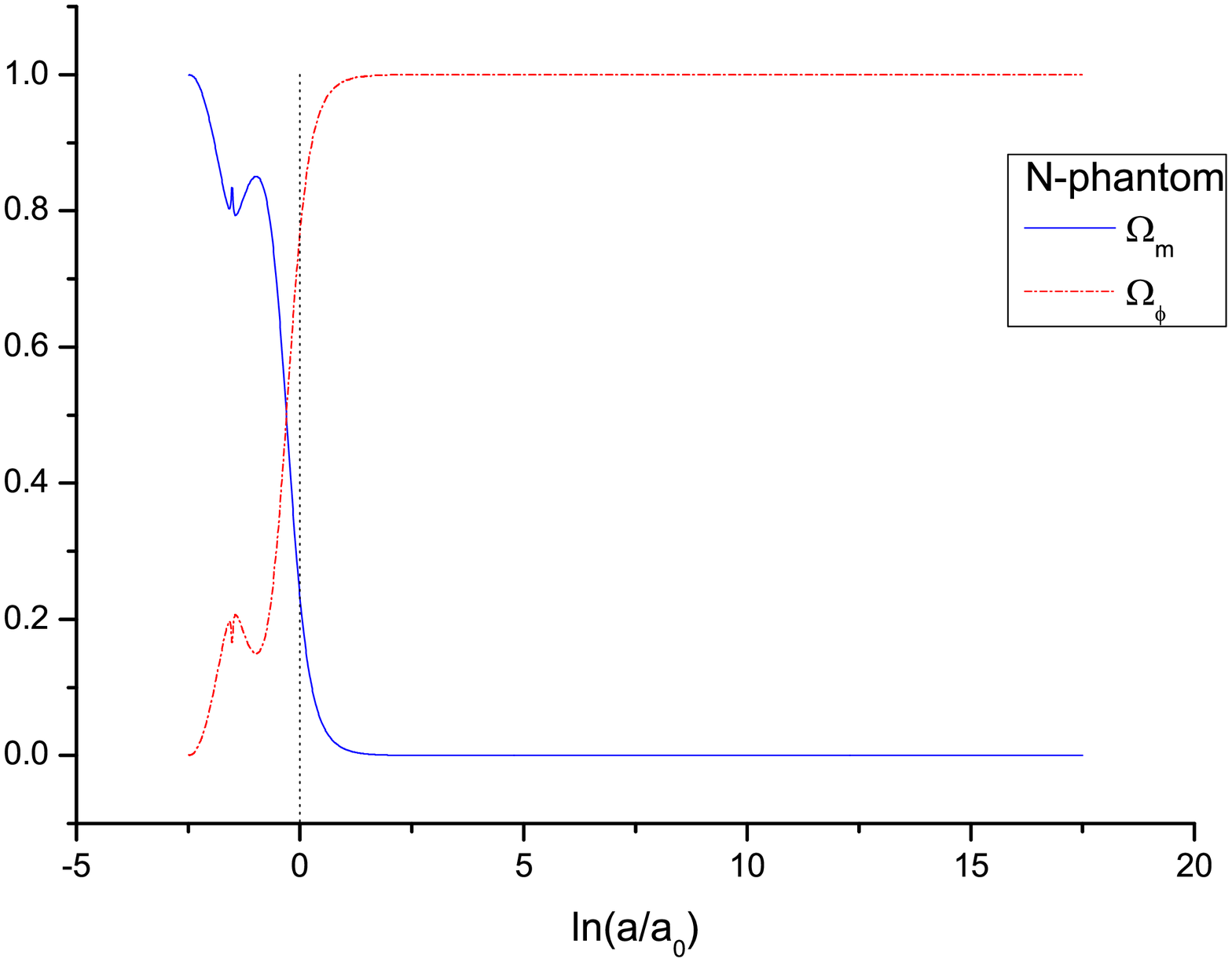}
\caption{\label{fig2} The evolutions of fractional energy densities
$\Omega_\phi$ and $\Omega_{m}$ in N-phantom model.}
\end{center}
\end{figure}


Based on the  exact potential forms  given by  Noether symmetry in
N-quintessence and N-phantom models, the evolution of our universe
could be analyzed. Firstly, two new variables $y=\phi/m_{pl}$,
$u=\ln(a/a_{0})$ are needed. Then we can define the fractional
energy density of dust matter as
$\Omega_{m}=\rho_m/3H^{2}m_{pl}^{2}=\Omega_{m0}(H_{0}/H)^{2}\exp(-3u)$,
and the fractional energy density of scalar fields
$\Omega_{\phi}=\rho_{\phi}/3H^{2}m_{pl}^{2}$ which  depends on the
exact potential form.

In  N-quintessence model, we discuss the possible accelerating
solution which is presented in  Eq.(\ref{V}) with $A=0$, $B\neq0$
and $\lambda=\sqrt{3/2}$.  $\Omega_{\phi}$  can be written down as
 \be
 \Omega_{\phi}=\frac{y'^{2}}{6}+\Omega_{V}\exp(-\lambda y),
 \ee
 where
 $\Omega_{V}=V_{0}B^{2}/3H^{2}m_{pl}^{2}=\Omega_{V0}(H_{0}/H)^{2}$.
Then, we can simplify Eq.(\ref{fr1}) and (\ref{me}) as
 \begin{eqnarray}
&&
(\frac{H}{H_{0}})^{2}=\frac{\Omega_{m0}\exp(-3u)+\Omega_{V0}\exp(-\lambda
y)}{1-y'^{2}/6},\\
 && y''=3\lambda\Omega_{V}\exp(-\lambda
y)-\left[\frac{3}{2}\Omega_{m}+3\Omega_{V}\exp(-\lambda y)\right]y',
\end{eqnarray}
where a prime denotes the derivative with respect to $u$. Following
the numerical calculation method used in
Ref.\cite{Capozziello:2009te,Gong:2006mn}, the evolution of the
fractional energy densities can be plotted. We choose $\Omega_{m}=1$
in  the matter dominated epoch around $a/a_{0}\approx1/12$ or
$u=-2.5$ as initial condition. Fig.\ref{fig1} shows today's
fractional density density $\Omega_{m0}$ is nearly $0.6$ which is
contradictable with the widest observational results
$\Omega_{m0}=0.3\pm 0.1$ \cite{Carlberg:1995aq}.

Furthermore,  setting $N=1$  and using  the potential in
Eq.(\ref{V2}), $\Omega_{\phi}$  in N-phantom case reads
 \be
 \Omega_{\phi}=\frac{y'^{2}}{6}+\Omega_{V}(1-\cos(\sqrt{\frac{3N}{2}}y)),
  \ee
  where
  $\Omega_{V}=V_{0}/3H^{2}m_{pl}^{2}=\Omega_{V0}(H_{0}/H)^{2}$.
Then, the  evolutions of scale
 factor and scalar field in N-phantom case are
\begin{eqnarray}
 &&
(\frac{H}{H_{0}})^{2}=\frac{\Omega_{m0}\exp(-3u)+\Omega_{V0}(1-\cos(\sqrt{\frac{3N}{2}}y))}{1-y'^{2}/6};\\
 &&
 y''=-\frac{3}{2}\lambda\Omega_{V}\sin(\sqrt{\frac{3N}{2}}y)-\left[\frac{3}{2}\Omega_{m}+\frac{3\Omega_{V}}{2}(1-\cos(\sqrt{\frac{3N}{2}}y))\right]y'.
\end{eqnarray}
 We can also start from  the matter
dominated epoch around $a/a_{0}\approx1/12$ or $u=-2.5$, and give
out the evolutions of the fractional energy densities.
Fig.\ref{fig2} shows  today's fractional density of dust matter
$\Omega_{m0}$ is nearly $0.23$ which is consistent with the
observational results $\Omega_{m0}=0.3\pm 0.1$.


\begin{figure}[ht]
\begin{center}
\includegraphics[width=15cm]{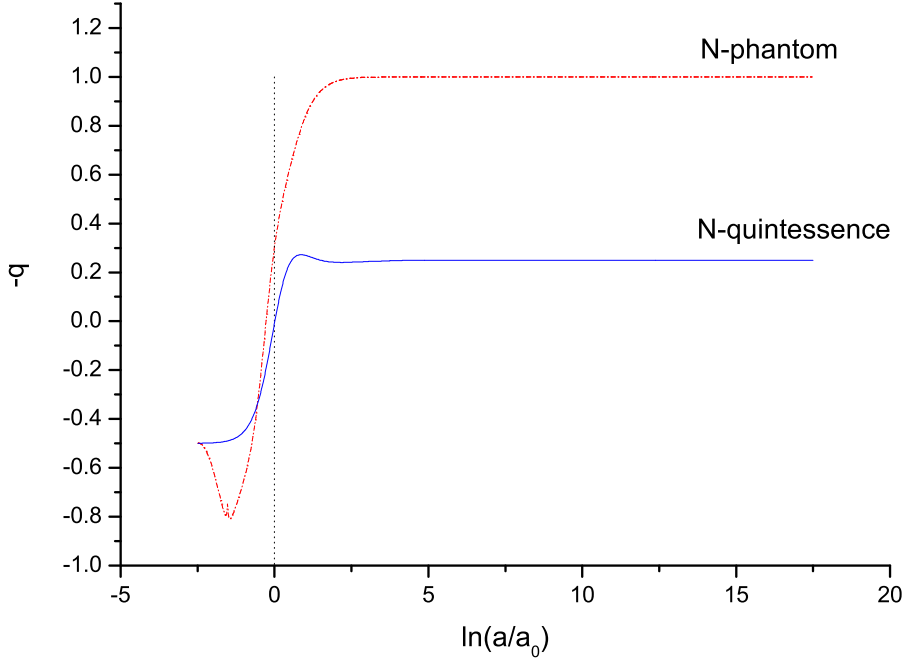}
\caption{\label{fig3} The evolutions of the minus of the
deceleration factor $-q$ in N-quintessence  and N-phantom cases.}
\end{center}
\end{figure}


Specifically speaking, we can write down the acceleration
 (the minus of the deceleration factor)
 \be
\frac{\ddot{a}}{aH^{2}}=-q=\Omega_{\phi}-\frac{1}{3}y'^{2}-\frac{1}{2}\Omega_{m},
 \ee
and plot its evolutions  in N-quintessence and N-phantom cases.
Fig.\ref{fig3} shows N-quintessence with exponential potential
chosen by Noether symmetry  cannot make our universe accelerate
($-q\leq0$), while  N-phantom with the cosine potential can make our
universe accelerate ($-q>0$). These results coincide with the
evolutions of the fractional energy densities in Fig.\ref{fig1} and
\ref{fig2}. In conclusion, today's acceleration heavily depends on
the choice of  potentials. This is the reason why we don't discuss
the evolution of our universe in the N-quintom case where the
Noether symmetry doesn't give out the exact form of potentials.

\section{Conclusion}\label{sec5}
There is no immediate physical justification for the choice of $V(\phi)$ in multiple scalar fields.
 In this Letter, to choose proper potentials for multiple scalar fields scenario,
 and to be consistent with the observations which indicates the EoS parameter in the range
 of $-1.21\leq \omega_{\phi}\leq -0.89$,
  we have studied the N-quintessence, N-phantom, N-quintom scalar fields models by the Noether symmetry approach.  The existence of Noether symmetry implies that  with
  respect to the infinitesimal generator of the desired symmetry, the Lie derivative of the related
Lagrangian vanishes.
As we have considered a flat FRW metric,
the phase space in the N-quintessence and N-phantom was then constructed by taking the scale factor $a$ and the scalar field $\phi$ as independent
dynamical variables. In the N-quintom case, we have to expand the configuration space to
$\mathcal{Q}=(a,\phi_{q},\phi_{p})$.

Specifically speaking, on the one hand, the Noether conditions
depend on the cosmological dynamics which is determined by the potentials. On the other hand,
the main consequence by adding the Noether symmetry  is that we have selected the class of potentials and indicated the most reasonable,
specific ones directly from the physical interpretation.
  In the N-quintessence case, we  find the exponential
potentials from the Noether conditions  which could be derived from
several theoretic models. In the  N-phantom case, the suitable
potential  required by the Noether conditions is
$\frac{V_{0}}{2}(1-\cos(\sqrt{\frac{3N}{2}}\frac{\phi}{m_{pl}}))$
which is related to pseudo Nambu-Goldstone boson. The case of the
N-quintom is very interesting. Although it does not give an explicit
potential, it gives a constraint on the forms of the scalar field
potentials.


\section*{Acknowledgements}
We thank Prof. Rong-gen Cai and Nan Liang for useful discussions. We
also thanks the anonymous referee for his/her useful suggestions.
This work was supported by CQUPT under Grant No.A2009-16, the
Ministry of Science and Technology of China national basic science
Program (973 Project) under grant Nos. 2007CB815401 and
2010CB833004, the National Natural Science Foundation of China key
project under grant Nos. 10533010 and 10935013, and the
Distinguished Young Scholar Grant 10825313, and the Natural Science
Foundation Project of CQ CSTC under Grant No. 2009BA4050.

\end{document}